\documentstyle[]{mn}
\input psfig.tex
\def\lsim{\lower.5ex\hbox{$\; \buildrel < \over \sim \;$}}
\def\gsim{\lower.5ex\hbox{$\; \buildrel > \over \sim \;$}}
\def \simeq{\lower.3ex\hbox{$\; \buildrel \sim \over - \;$}}
\def\ch{\lower-0.55ex\hbox{--}\kern-0.55em{\lower0.15ex\hbox{$h$}}}
\def\lh{\lower-0.55ex\hbox{--}\kern-0.55em{\lower0.15ex\hbox{$\lambda$}}}
\newif\ifAMStwofonts
\ifoldfss
  \ifCUPmtlplainloaded \else
    \NewTextAlphabet{textbfit} {cmbxti10} {}
    \NewTextAlphabet{textbfss} {cmssbx10} {}
    \NewMathAlphabet{mathbfit} {cmbxti10} {} 
    \NewMathAlphabet{mathbfss} {cmssbx10} {} 
  \fi
  \ifAMStwofonts
    \ifCUPmtlplainloaded \else
      \NewSymbolFont{upmath} {eurm10}
      \NewSymbolFont{AMSa} {msam10}
      \NewMathSymbol{\upi}     {0}{upmath}{19}
      \NewMathSymbol{\umu}     {0}{upmath}{16}
      \NewMathSymbol{\upartial}{0}{upmath}{40}
      \NewMathSymbol{\leqslant}{3}{AMSa}{36}
      \NewMathSymbol{\geqslant}{3}{AMSa}{3E}

    \fi
  \fi
\fi 
\ifnfssone
  \newmathalphabet{\mathit}
  \addtoversion{normal}{\mathit}{cmr}{m}{it}
  \addtoversion{bold}{\mathit}{cmr}{bx}{it}
  \newmathalphabet{\mathbfit} 
  \addtoversion{normal}{\mathbfit}{cmr}{bx}{it}
  \addtoversion{bold}{\mathbfit}{cmr}{bx}{it}
  \newmathalphabet{\mathbfss} 
  \addtoversion{normal}{\mathbfss}{cmss}{bx}{n}
  \addtoversion{bold}{\mathbfss}{cmss}{bx}{n}
  \ifAMStwofonts
    \ifCUPmtlplainloaded \else
      \UseAMStwoboldmath
      \makeatletter
      \new@mathgroup\upmath@group
      \define@mathgroup\mv@normal\upmath@group{eur}{m}{n}
      \define@mathgroup\mv@bold\upmath@group{eur}{b}{n}
      \edef\UPM{\hexnumber\upmath@group}
      \new@mathgroup\amsa@group
      \define@mathgroup\mv@normal\amsa@group{msa}{m}{n}
      \define@mathgroup\mv@bold\amsa@group{msa}{m}{n}
      \edef\AMSa{\hexnumber\amsa@group}
      \makeatother
      \mathchardef\upi="0\UPM19
      \mathchardef\umu="0\UPM16
      \mathchardef\upartial="0\UPM40
      \mathchardef\leqslant="3\AMSa36
      \mathchardef\geqslant="3\AMSa3E
    \fi
  \fi
\fi 

\ifnfsstwo
  \DeclareMathAlphabet{\mathbfit}{OT1}{cmr}{bx}{it}
  \SetMathAlphabet\mathbfit{bold}{OT1}{cmr}{bx}{it}
  \DeclareMathAlphabet{\mathbfss}{OT1}{cmss}{bx}{n}
  \SetMathAlphabet\mathbfss{bold}{OT1}{cmss}{bx}{n}
  \ifAMStwofonts
    \ifCUPmtlplainloaded \else
      \DeclareSymbolFont{UPM}{U}{eur}{m}{n}
      \SetSymbolFont{UPM}{bold}{U}{eur}{b}{n}
      \DeclareSymbolFont{AMSa}{U}{msa}{m}{n}
      \DeclareMathSymbol{\upi}{0}{UPM}{"19}
      \DeclareMathSymbol{\umu}{0}{UPM}{"16}
      \DeclareMathSymbol{\upartial}{0}{UPM}{"40}
      \DeclareMathSymbol{\leqslant}{3}{AMSa}{"36}
      \DeclareMathSymbol{\geqslant}{3}{AMSa}{"3E}
    \fi
  \fi
\fi 

\ifCUPmtlplainloaded \else
  \ifAMStwofonts \else 
    \def\upi{\pi}
    \def\umu{\mu}
    \def\upartial{\partial}
  \fi
\fi

\title{On the source of QPO of the black hole candidate GRS1915+105: some new observations
and their interpretation.}

\author[Anuj Nandi, Sivakumar G. Manickam, A.R. Rao and Sandip K. Chakrabarti]
       {Anuj Nandi$^1$, Sivakumar G. Manickam$^1$, A.R. Rao$^2$ and Sandip K. Chakrabarti$^{1,3}$\\
$^1$ S.N. Bose National Centre for Basic Sciences,\\
JD-Block, Sector III, Salt Lake, Calcutta 700098, India;\\
$^2$ Tata Institute of Fundamental Research, Homi Bhabha Road, Colaba, Mumbai 400005\\
$^3$ Centre for Space Physics, 114/v/1A Raja S.C. Mullick Rd., Calcutta 700047, India\\
anuj@boson.bose.res.in, sivman@boson.bose.res.in, arrao@tifr.res.in, chakraba@boson.bose.res.in}

\date{Accepted .
      Received ;
      in original form }
\pubyear{2000}
\begin{document}

\maketitle

\begin{abstract}

A few classes of the light curve of the black hole 
candidate GRS 1915+105 have been analyzed in detail. 
We discover that unlike the previous findings, QPOs
occasionally occur even in the so-called `On' or softer states. 
Such findings may require a revision 
of the accretion/wind scenario of the black hole 
candidates. We conjecture that considerable winds which is produced
in `Off' states,  cool down due to Comptonization, and
falls back to the disk and creating an excess accretion rate to produce the 
so-called `On' state. After the drainage of the excess matter,
the disk goes back to the `Off' state.  Our findings strengthen the 
shock oscillation model for QPOs. 

\end{abstract}


\noindent MNRAS (to appear)

\section{Introduction}

The black hole candidate GRS1915+105 continues to excite astrophysicists by 
having one of the most, if not the most mysterious light curves. In a matter 
of days, the light curve changes its character, and within each day, 
photon counts show variations of a factor of two to five or more
(Morgan, Remillard \&  Greiner, 1997; Belloni et al, 1997; Paul et al. 
1998; Manickam \& Chakrabarti 1999). The power density spectra 
shows clear evidence of quasi-periodic oscillations (QPOs) with frequency 
ranging from $\sim 0.001$Hz to $67$Hz (see, e.g., Morgan, Remillard \&  
Greiner, 1997; Chakrabarti \& Manickam 2000, hereafter CM00, and references therein).

The origin of the quasi-periodic oscillations cannot be a 
complete mystery since there are very clear evidence that 
sub-Keplerian flows, which must occur in a {\it black hole} accretion, 
exhibit very wild time-dependent behaviour including large amplitude 
shock oscillations. This is true, especially when the infall time 
matches with the cooling time (Molteni, Sponholz \& Chakrabarti, 
1996; Ryu, Chakrabarti \& Molteni 1997; Paul et al, 1998; 
Remillard et al, 1999ab; Muno, Morgan \& Remillard, 1999).
This became more evident when it was observed that the 
QPOs are absent in low energy photons but is very strong
at high energies, supporting the view that the photons participating 
in QPOs originate at hot post-shock flow (CM00, Rao et al. 2000). 
It has also been observed that QPOs are seen mostly when 
the photon counts are low (`Off' states) and QPO is very 
weak or absent in the On states when the 
photon counts are about two  to  five times larger. In the present
{\it Letter}, by On and Off states, we would mean the high and low count states
of the light curves. This nomenclature is not related to spectral states.

Because of the presence of the On and the Off states in the light curves, 
explanation of QPOs using radial oscillation of the centrifugal pressure supported shocks
alone (which produce the right amplitude and frequency for the QPOs)
cannot be the complete story. The very fact that the Off states 
terminate and the On states emerge in almost regular basis (but not exactly regular)
gives rise to another time scale which must, at the same time,
be dependent on disk/jet parameters. This is because Belloni 
et al. (2000) found that at least twelve types of light 
curves are seen, and within each type, the duration 
and behaviour of the On and the Off (if both exist) states were not at all 
fixed. For instance, $\rho$ class 
exhibits extremely regular light curves (Taam, Chen \& Swank, 1997; Vilhu \& Nevalainen, 1998)
with broad Off- or low-count states and very narrow, spiky, high-count or On-states (see, Fig. 2b below). 
Light curves in the $\nu$ class is similar to those of $\rho$, but are highly irregular (see, Fig. 2a below).
In $\lambda$ class, both Off and On states are of longer time duration (Belloni et al. 1997) 
while in $\kappa$ class these durations are relatively shorter (see, Fig. 5 below).
Nandi, Manickam and Chakrabarti (2000)
using a completely different procedure divided the 
light curves into four fundamental types (Hard, Soft, Semi-soft 
and Intermediate). The Intermediate class shows On/Off transitions.
Chakrabarti (1999) and Chakrabarti \& Manickam (2000) 
showed that outflow rates from the centrifugal barrier must play 
a major role in deciding the duration, as the wind matter 
at least up to the sonic sphere can be Comptonized 
and cooled down. Most of this
cold matter (below the sonic surface of the cooler wind) can 
fall back on the immediate vicinity of the disk increasing 
its accretion rate temporarily while the rest must separate 
out at a supersonic speed. As this excess matter drains out, 
the Off state together with the QPO start appearing again. 

In this {\it Letter},  we analyze light curves in detail 
and find that QPOs may be observed at very unlikely times
in the light curves. For instance, we find that 
very often, there is a sharp peak (`first hiccup') 
at the onset of the On state and there is a 
sharp peak (`last hiccup') just prior to going to the Off 
state. The first peak, though in the On state, very often shows 
QPOs while the last peak, though the radiation is much 
harder, does not show a QPO. We also note that the
radiation progressively hardens in the On state. 
In several cases, using data from both the RXTE and Indian X-ray Astronomical Experiment (IXAE),
$\rho$ class curves (mostly mini-$\rho$ type) are seen to be peeled off from a $\kappa$ class.  
In $\kappa$ and $\lambda$, the On state duration is long and 
just before going to the Off state, the light curve becomes 
noisy and oscillating in nature, and the features indicate as though
the light curve is made up of `sums' of $\rho$ types. Only the 
lower half of the oscillations (local Off states of mini-$\rho$ states) exhibit QPO!
We believe that these observations definitely point to the drainage of 
extra matter in the disk which was accumulated from the wind.

One of the problems in analyzing data of one of the most complex 
objects, such as GRS1915+105 is that one has to ask the right questions
and as many of them as possible. Once a paradigm is kept in the back of the
mind, asking questions become easier. We therefore concentrate on 
models which required sub-Keplerian and Keplerian flows simultaneously
and where winds are also produced self-consistently. In the next Section, we
present some of the `subtle' observational results which have not been reported by
workers before (despite the fact that GRS 1915+105 is probably the 
most studied object in recent years). In \S 3, we interpret this observations
in terms of the advective disk paradigm. In \S 4, we draw our conclusions.

\section{Observations}

\subsection{Properties of two peaks in the On state}

GRS 1915+105 exhibits both broad and narrow On states (Paul et al. 1998;
Yadav et al, 1999; Belloni et al. 2000; CM00). When the On state is narrow 
or spiky, i.e., the duration with high photon counts is very small, and 
very often, just after transition from On to Off state, a second peak 
is observed. This has also been cursorily reported by Paul et al. (1998).
Figure 1 shows details of a narrow section of the all-channel
light curve corresponding to the RXTE observation of 15th of Oct. 
1996 (The observation ID number is 10408-01-41-00.). We denote 
the peak appearing first as the primary peak (P1 for short) and the peak following P1
as the secondary peak (P2 for short) respectively. The words Off, `P1' and `P2' are marked. 
The entire light curve is almost a repetition of this section. Figure 2a 
presents the light curves of this data clearly showing two peaks P1 and P2 in the On state.
The time lag between the two peaks is roughly constant on a given day 
(in this case about $10$ seconds) and we do not find any correlation 
between this lag and the duration of the Off state. The light 
curve is shown in four panels: the panels are drawn for channel 
energies (from top to bottom) $0-5.07$keV, $5.43-6.88$keV, 
$7.24-9.43$keV, and $9.79-13.09$keV respectively (corresponding 
channels are $0-13$, $14-18$, $19-25$, $26-35$ respectively).
Note that while in low energies, photon counts in P1 is much larger
compared to P2, in higher energies they are roughly equal, suggesting that 
the spectrum of P2 is harder. Taam, Chen and Swank (1997)
also noted the existence of these peaks and that P2 is harder compared
to P1. A similar light curve is shown in Fig. 2b, 
for the RXTE observation of 22nd of June, 1997
(see also Vilhu \& Nevalainen (1998) and Yadav et al (1999)
for displaying this light curve.). The corresponding 
observation ID is 20402-01-34-01. In this case, the time lag 
is about $4$ seconds but other features are similar to those 
of Fig. 2a. The channel energies are marked on each panel 
(corresponding channels numbers are $0-13$, $14-19$, $20-25$, $26-35$ 
respectively). 

\begin {figure}
\vbox{
\vskip -1.0cm
\hskip 0.0cm
\centerline{
\psfig{figure=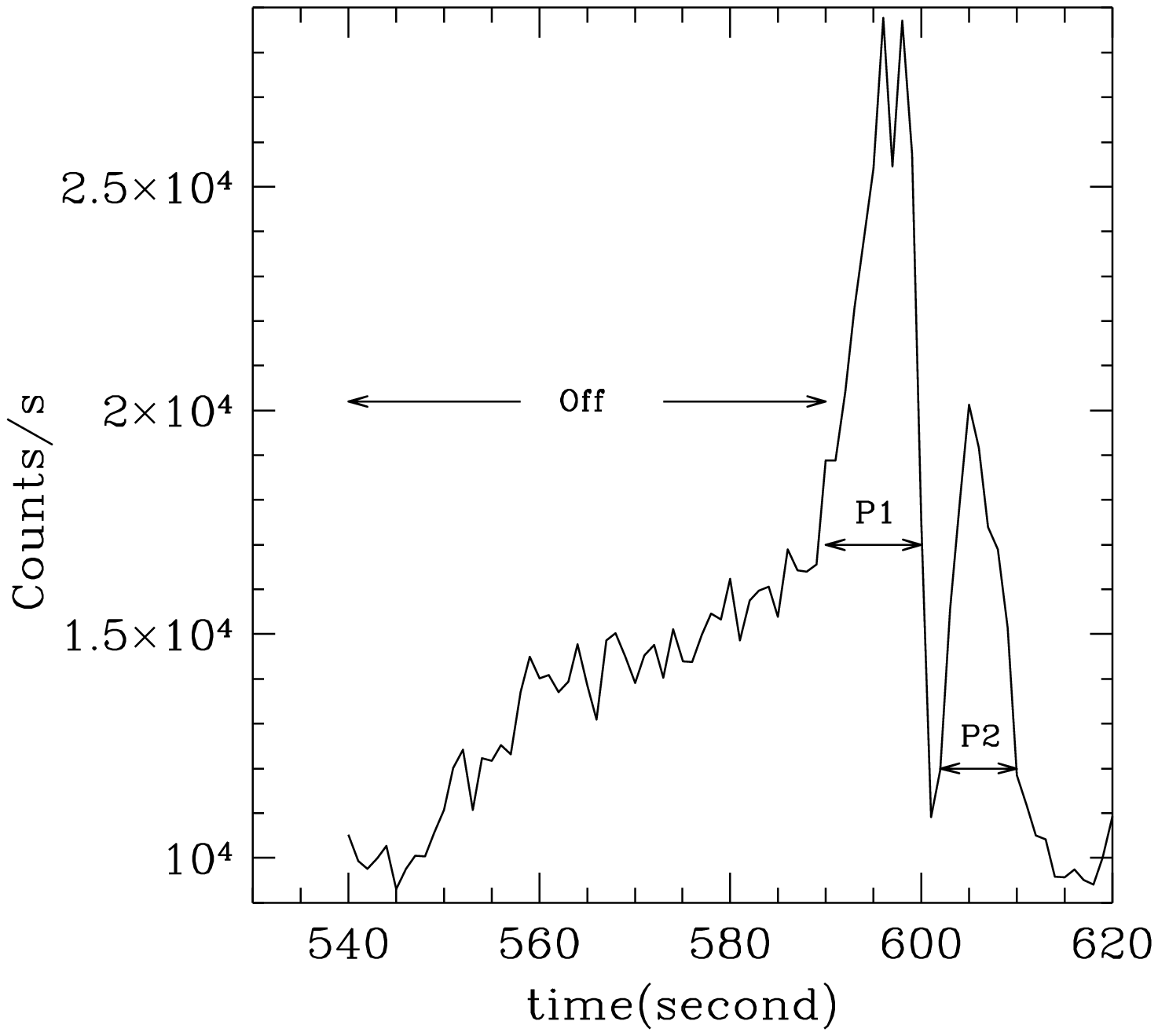,height=10truecm,width=10truecm,angle=0}}}
\vspace{-2.0cm}
\noindent {\small {\bf Fig. 1}: 
Subdivisions of a quasi-repetitive structure of the light 
curve of Oct. 15th, 1996 (ID:10408-01-41-00). Relatively hard
radiations are emitted in the region marked Off with relatively lower count rates. 
There are typically two major peaks, P1 followed by P2. }
\end{figure}

\begin {figure}
\vbox{
\vskip -1.0cm
\hskip 1.0cm
\psfig{figure=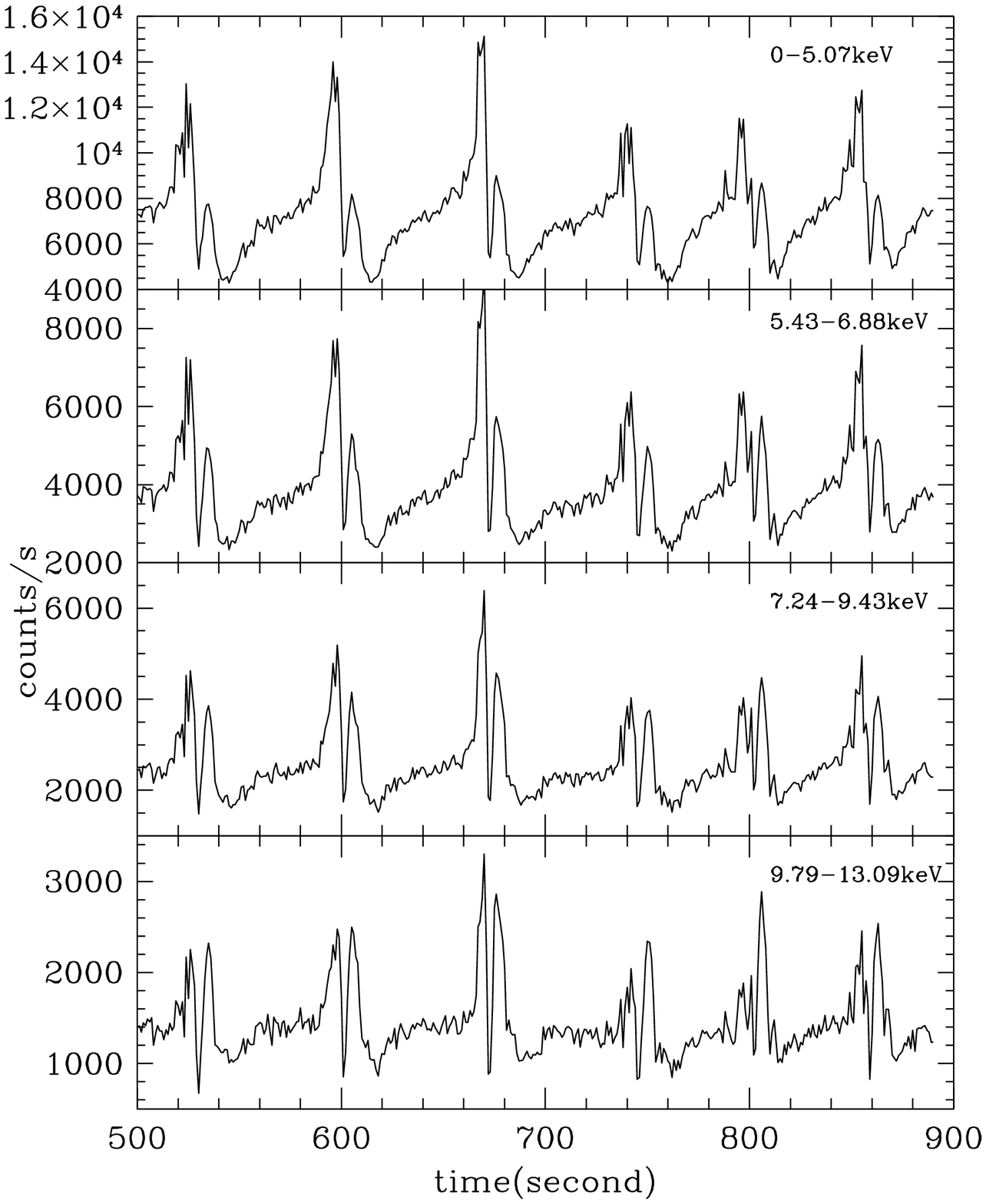,height=10truecm,width=10truecm,angle=0}}
\vspace{-0.5cm}
\noindent {\small {\bf Fig. 2a}: Four panels showing light curves of a 
part of the observation on Oct. 15th, 1996 (ID: 10408-01-41-00) at different channels (energies 
are marked). Note that ratio of photon counts in P1 and P2 tends to become unity at higher energies.}
\end{figure}

\begin {figure}
\vbox{
\vskip -1.0cm
\hskip 1.0cm
\psfig{figure=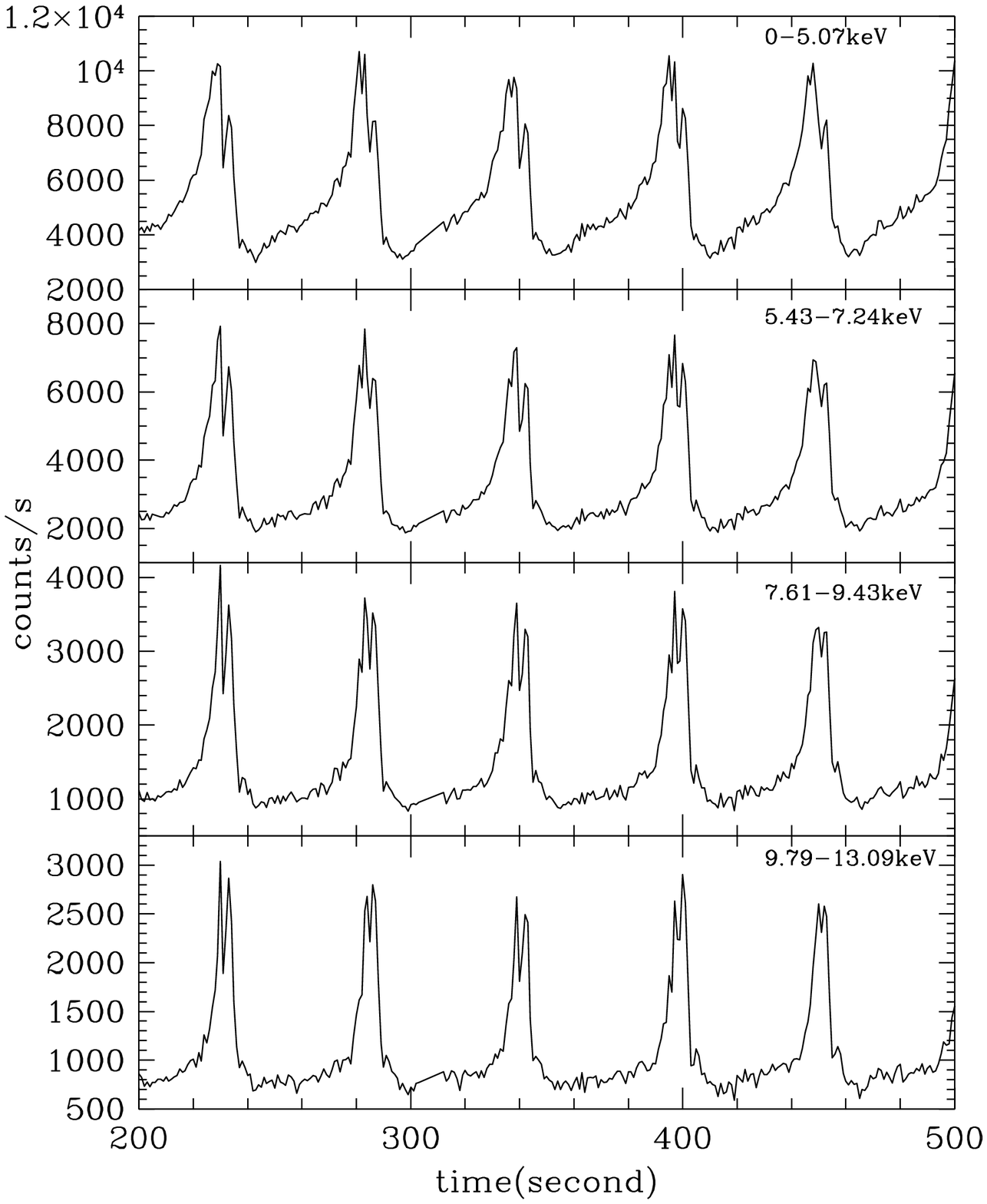,height=10truecm,width=10truecm,angle=0}}
\vspace{-0.5cm}
\noindent {\small {\bf Fig. 2b} :
Four panels showing light curves of a part of the observation on June 22nd, 1997 (ID: 20402-01-34-01)
at different channels (energies are marked). }
\end{figure}

\begin {figure}
\vbox{
\vskip 0.0cm
\hskip 0.0cm
\centerline{
\psfig{figure=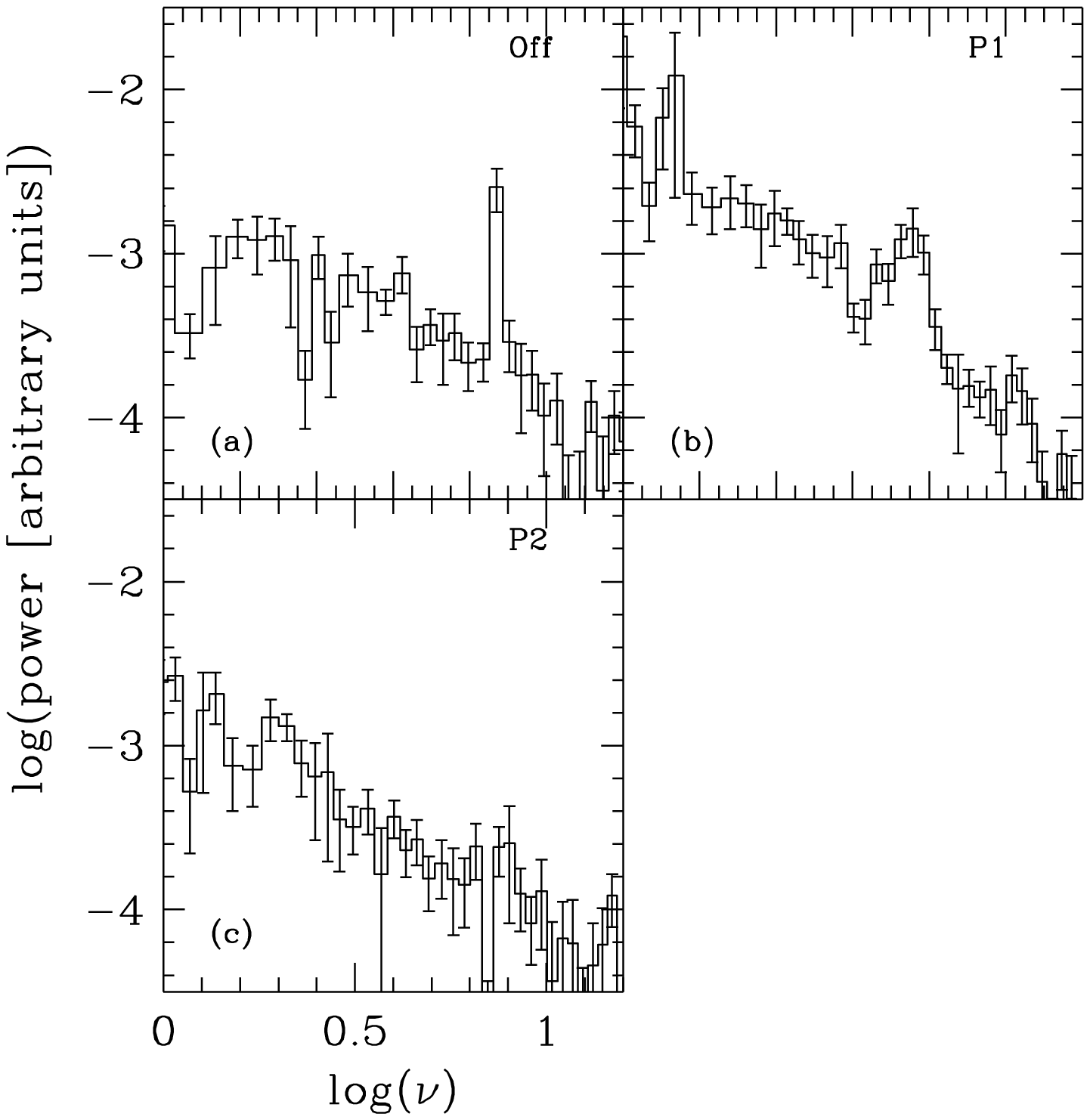,height=10truecm,width=10truecm,angle=0}}}
\vspace{-1.0cm}
\noindent {\small {\bf Fig. 3(a-c)}: Power Density Spectrum (PDS) of three regions 
marked in Fig. 1 of the observation dated Oct. 15th, 1996. Several data segments
have been added to improve statistics. Note the absence of QPO in P2. }
\end{figure}

\begin {figure}
\vbox{
\vskip 0.0cm
\hskip 0.0cm
\psfig{figure=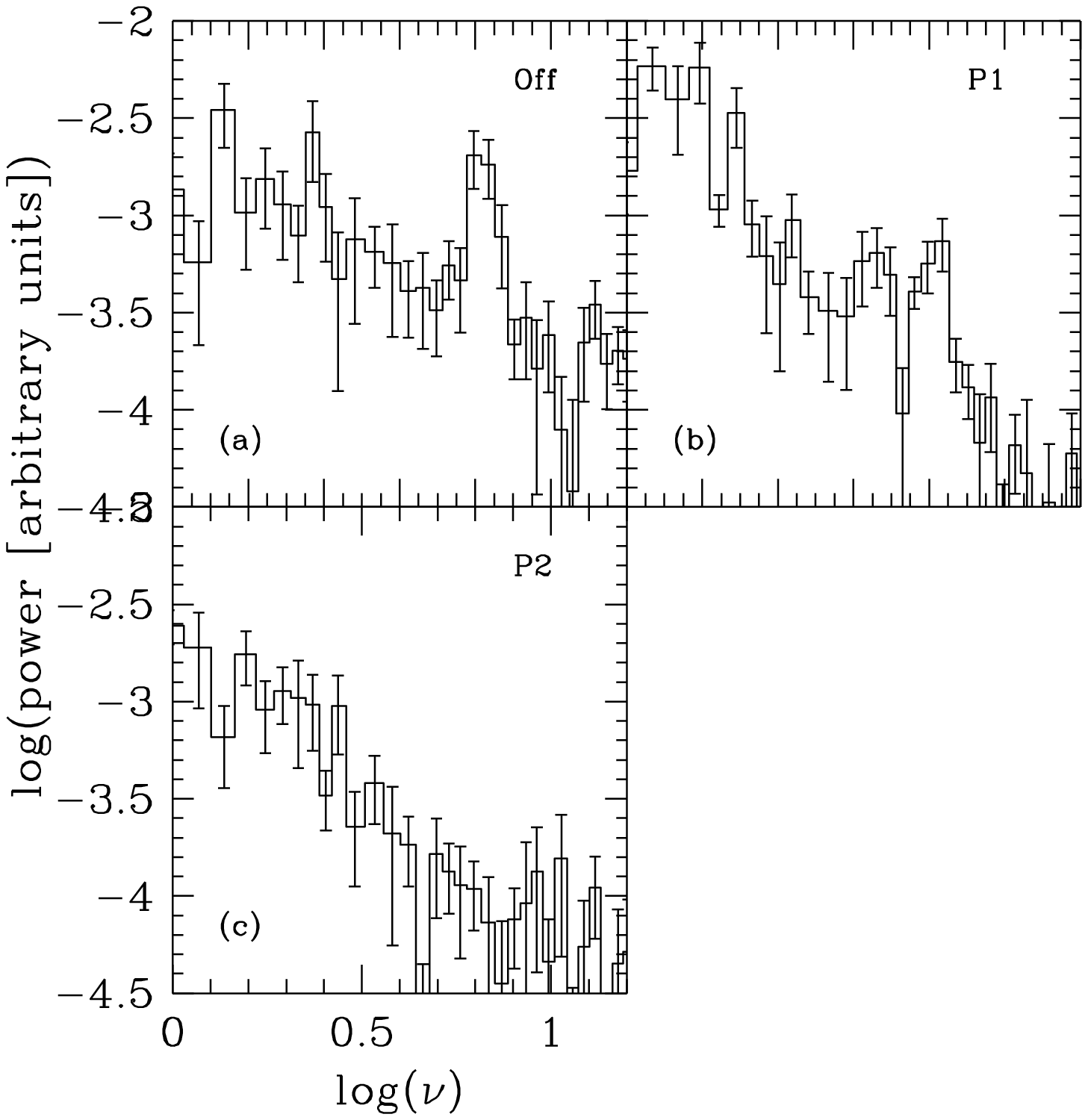,height=10truecm,width=10truecm,angle=0}}
\vspace{-3.0cm}
\noindent {\small {\bf Fig. 4(a-c)}: Power Density Spectrum (PDS) of three regions 
marked in Fig. 1 of the observation dated June 22nd, 1997. Several data segments
have been added to improve statistics. Note the absence of QPO in P2. }
\end{figure}

Figures 3(a-c) show the power density spectra (PDS) of the three regions marked in Fig. 1.
While selecting photons from `P1' and `P2' regions we took special care that 
they are not contaminated by the photons from the  Off states. Also, to improve 
statistics, we added data from many peaks over the entire duration of the 
observation on that day. Note that in the Off state there is a distinct QPO 
of frequency $\nu_{qpo}=7.4$Hz. Photons in `P1' also exhibit QPO though it 
is weaker ($\nu_{qpo}\sim 5.7$Hz). QPO is completely absent from `P2'. We examined 
the RXTE data of 22nd June, 1997 also, the PDS of which is shown in Fig. 4(a-c). 
The result is generally the same. The QPO frequency in the Off-state is 
given by $\nu_{qpo}=6.3$Hz and in P1 peak $\nu_{qpo}=6.8$Hz.  We therefore 
believe that the observed features may be generic. 

\subsection{Properties of On$^{++}$ state}

It is generally observed that whenever the duration 
of On or the high count state is large, light curves 
become very noisy and count rates start oscillating wildly 
as the Off state is approached. The later half may be termed 
as `On$^{++}$' state and Manickam \& Chakrabarti (1999) showed 
that this region exhibits a weak QPO. In fact, similar to Fig. 
2(a-b) above, where the photons in P2 are harder compared 
to P1, one finds that On$^{++}$ state is harder compared to the
first half of the On state. This is demonstrated in Fig. 5, 
where a part of the light curve from the so-called $\kappa$-state 
is shown. The energy ranges are marked in each panel. 
Upper four channels are from RXTE and the lower two channels
are from the Indian X-ray Astronomy Experiment (IXAE) data. The observation ID 
for RXTE is 20186-03-01-02. In both the cases one second bin-size 
is chosen. 

\begin {figure}
\vbox{
\vskip -3.0cm
\hskip 0.0cm
\psfig{figure=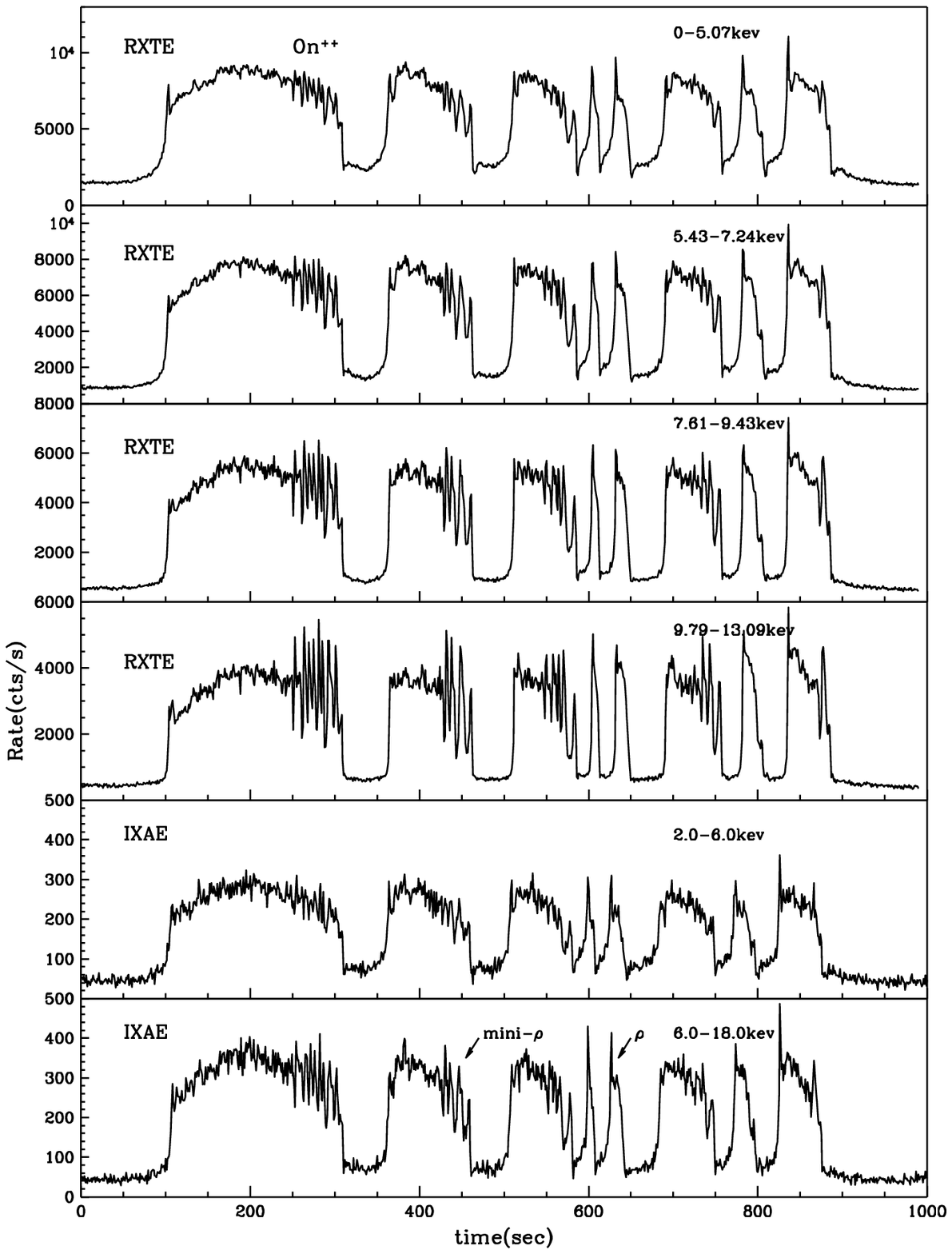,height=12truecm,width=12truecm,angle=0}}
\vspace{0.0cm}
\noindent {\small {\bf Fig. 5}: 
A part of the light curve from the so-called $\kappa$-state 
is shown. The energy ranges are marked in each panel. 
Upper four panels are from RXTE and the lower two panels
are from IXAE data. Note that in On$^{++}$ states light curves are
more noisy and oscillatory, more so for photons at higher energy.
A small arrow indicates a $\rho$ type burst.
Presence of mini-$\rho$ bursts within a $\kappa$ type On states 
indicates that $\rho$ may be the {\it fundamental} type of all burst.}
\end{figure}

First note that RXTE and IXAE show very similar behaviours 
throughout the period of overlap of observations. Second, 
towards the end of {\it each} of the On states (which we 
term as On$^{++}$ state for brevity), the light curves 
are noisy and generally oscillatory in nature. Third, in 
both the experiments, the relative oscillations are 
increased with the increasing photon energy in On$^{++}$
state. Fourth, though generally the light curve may be called 
that of a $\kappa$-class (Belloni et al. 2000), several pieces of $\rho$
class is evident. In fact, mini-$\rho$ type light 
curves are also evident in the On$^{++}$ states giving clear
evidence that the light curve in the $\rho$-class is more 
primitive. In the next Section we discuss possible 
interpretations of this important observation. 

In Fig. 6, we show a part of the light curve on the same day,
and draw two boxes (in dotted curves) at the On$^{++}$ state.
The photons from the upper box show no sign of QPO, while the
photons from the lower box shows clear evidence of QPO. Since
lower box contains photons which are from mini-Off states
of the mini-$\rho$ class mentioned above, it is not surprising
that this show QPOs. Indeed, while the Off state on this 
day shows QPO of frequency $3.09$Hz, these mini-Off states
show QPOs of frequency $6.25$Hz. This indicates that towards the
end, the shock again starts developing much closer to the black hole
giving rise to a higher oscillation frequency and when the shock is fully 
developed, the Off state begins with a QPO at $3.09$Hz.   

An easily missed phenomenon in all these light curves is that
most of the On states begin with a `hiccup' or a small peak, 
which may be likened with `P1' of $\rho$ or $\nu$ class and 
also end with another `hiccup' which may be likened with `P2'.  

\section{Possible interpretations of these observations}

\subsection {The Paradigm}

While there is as yet no fully self-consistent model
which includes disks, winds and radiative transfer simultaneously,
one can collect bits and pieces of the solutions and construct a
viable model for the system. Chakrabarti (1996) and more 
recently Chakrabarti (2000a) have discussed such solutions in detail.
Fig. 7 shows a cartoon diagram of an accretion/wind system of 
GRS 1915+105. Generally, it is assumed the the accretion is
advective in nature and has a centrifugal barrier dominated 
region which may or may not have become a fully developed shock (at $r=r_s$)
throughout the period of observation. Chakrabarti (1999)
showed that the centrifugal barrier dominated boundary layer (CENBOL for short) is not 
only responsible for the Comptonized radiation, but also  
responsible for the wind/jet formation. A very simple analysis
which envisages an isothermal wind at least up to the sonic surface (at $r=r_c$)
shows that winds should not be emitted in soft states, and 
very hard states should have very little winds. If the shock 
is of intermediate strength, outflow rate is very much high.

If the outflow rate is high enough, it can fill the sonic sphere (of size
$r_c \sim 2-3 r_s$; sub-sonic region up to the sonic surface) rapidly till the optical 
depth due to Compton scattering become larger than unity. Comptonization
cools down this region rapidly. CM00 suggests that the 
duration of the Off state is the time in which this region 
achieves this threshold of cooling. Since the specific energy and
angular momentum of the flow decide shock location and its strength
(Chakrabarti 1989), a small variation of the overall viscosity would
change the shock location, and therefore cause the variation of the
duration and the QPO frequency from one Off state to another. 
The correlation between duration and frequency based on this
consideration has been discussed in CM00. An important bye-product 
of all these is that as the disk loses matter (and pressure) from the post-shock region 
in the form of outflows, the shock, as well as the inner edge of the 
Keplerian disk moves inwards, gradually increasing the QPO 
frequency. This has been demonstrated by the dynamic power 
density spectra (e.g., Muno et al., 2000; Trudolyubov et al. 2000).

\begin {figure}
\vbox{
\vskip -1.0cm
\hskip 0.0cm
\psfig{figure=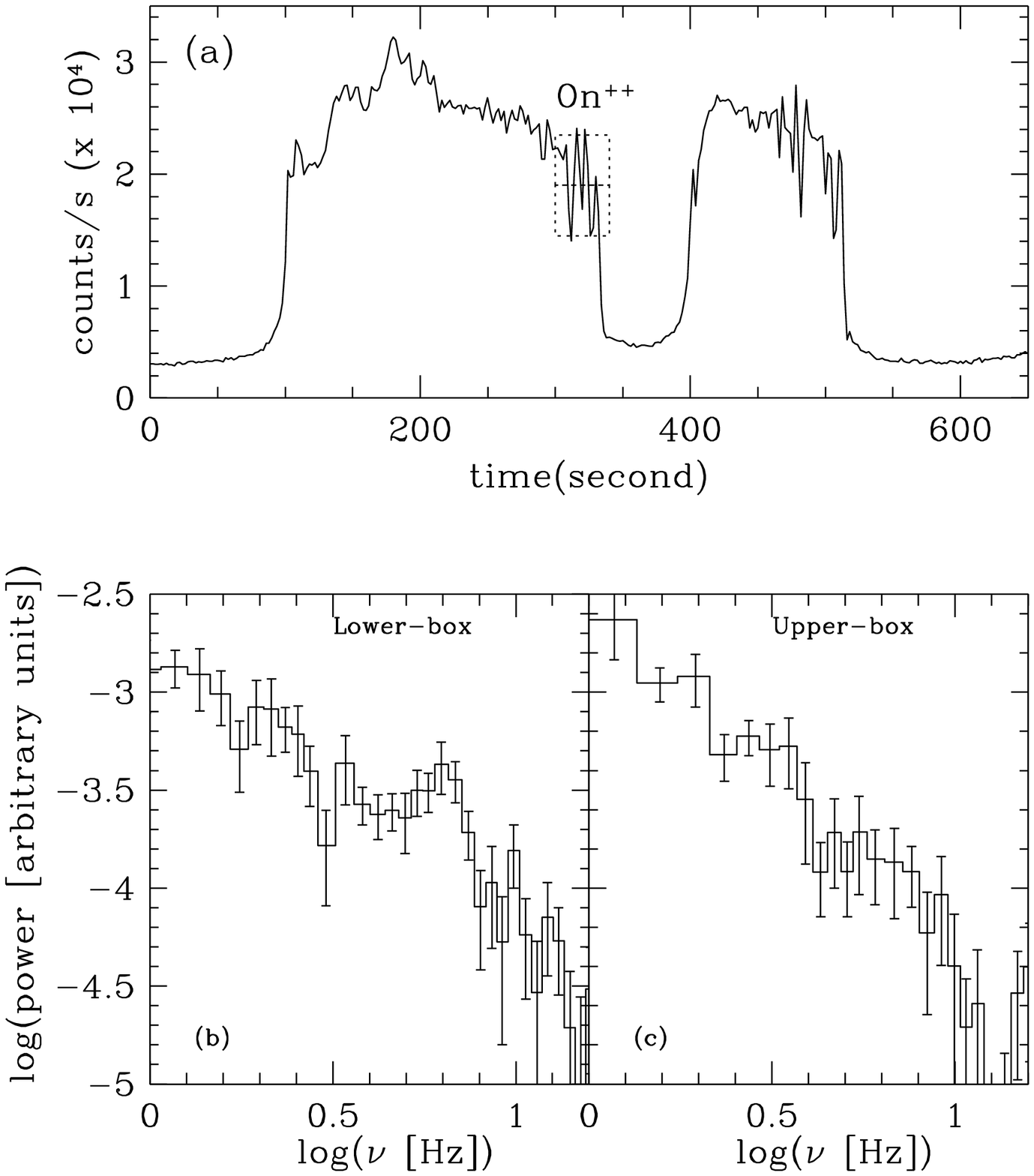,height=10truecm,width=10truecm,angle=0}}
\vspace{-1.0cm}
\noindent {\small {\bf Fig. 6}: 
A part of the light curve from the so-called $\kappa$-state 
is shown. In On$^{++}$ states, two dotted boxes are drawn and below
the PDS of each box is plotted. Note the presence of QPO in
lower box, which are nothing but mini-Off states
corresponding to mini-$\rho$ bursts.}
\end{figure}

A part of the wind which does not reach escape velocity 
must return back to the disk and the accretion rate of the
disk is temporarily modulated. This has been demonstrated by 
two dimensional simulations of advective disks (Molteni, 
Lanzafame \& Chakrabarti, 1994; Molteni, 
Ryu and Chakrabarti, 1996 and Ryu, Chakrabarti  \& Molteni, 1997).
This feedback mechanism becomes more complex in presence of 
radiative cooling of the outflow. After the wind is cooled 
down, the sound speed is reduced, and location of the 
sonic surface comes closer to the black hole, very abruptly.
Part of the originally subsonic outflowing matter still remains
below the new sonic surface and falls back since it loses drive to
escape. The rest becomes supersonic (because of reduced sound speed)
and separates in the form of blobs.
Thus, blobs are produced at the end of the off-states. This conjecture 
has been tested by multiwavelength observations (Dhawan et al. 2000). 

During the On states, the shock and the CENBOL are non-existent and hence the QPOs
are generally absent. Towards the end of the On state, in 
the so-called On$^{++}$ state, the shocks start forming close 
to the black hole because of heat generated by excess accretion. 
The shock then rapidly recedes to a distance consistent
with the steady state solution as soon as the excess matter is drained 
out of the disk. This causes the onset of the Off state.  This process of 
receding are regularly seen in numerical simulation of sub-Keplerian flows
(Chakrabarti \& Molteni, 1993; MLC94). 

\subsection{Origin of hiccups or P1 and P2 peaks}

\begin {figure}
\vbox{
\vskip 0.0cm
\hskip 0.0cm
\psfig{figure=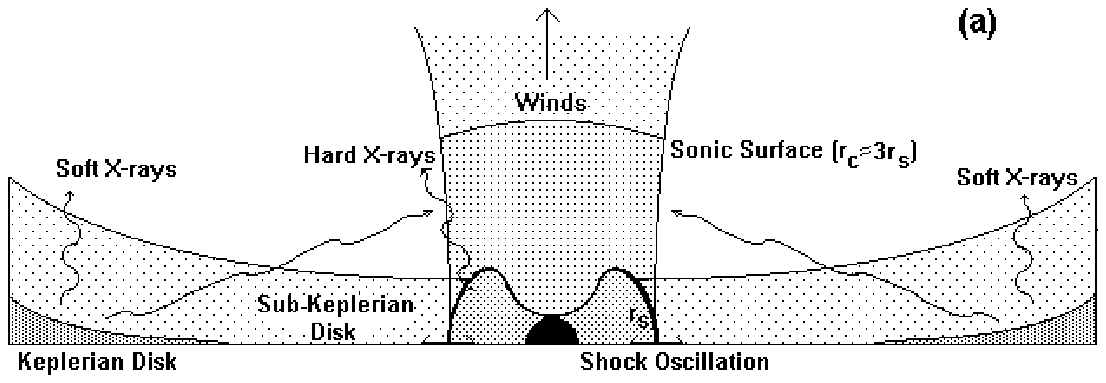,height=6truecm,width=8truecm,angle=0}
\vspace{2.0cm}
\psfig{figure=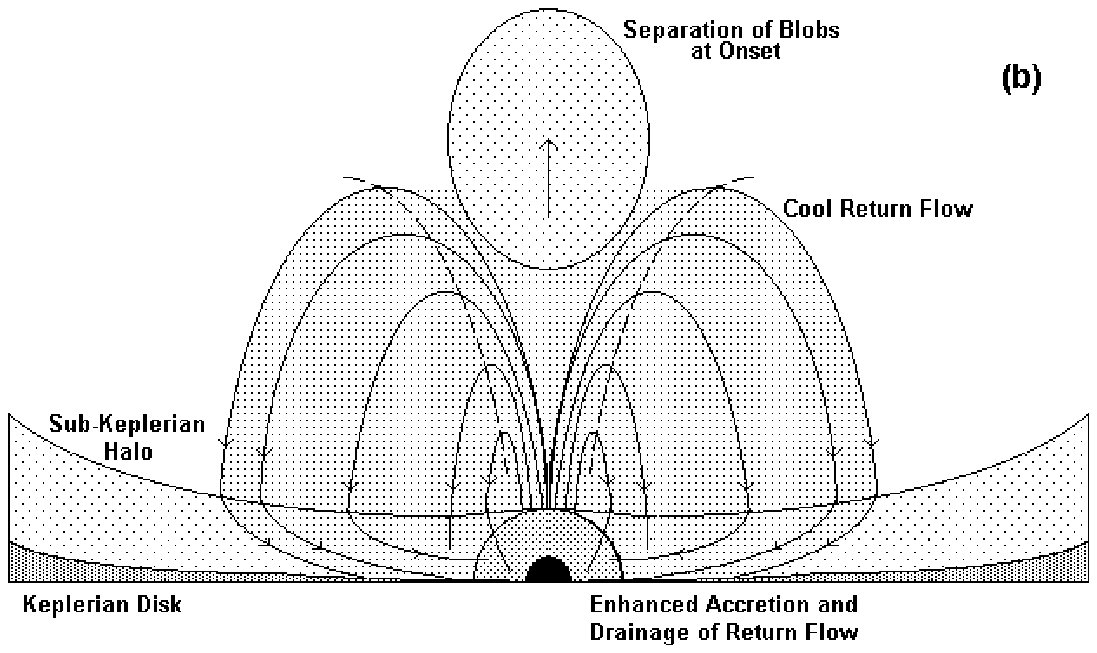,height=6truecm,width=8truecm,angle=0}}
\vspace{-1.0cm}
\noindent {\small {\bf Fig. 7}: 
Cartoon diagram showing possible physical processes close to a black hole
during an average time of the (a) Off and (b) On states. Centrifugal barrier or 
CENBOL produces winds. Subsonic region is cooled and falls back on the disk while
supersonic region separates out as blobs. Excess accretion inner part
causes higher counts in On states.}
\end{figure}

It is clear that since these peaks are separated by a few seconds,
and generally one has a QPO while the other does not, their origins
cannot be the same. From the paradigm described above, one may imagine 
that P1 forms when catastrophic cooling of the CENBOL-sonic sphere system
takes place. Since P1 is a continuation of the Off state, QPO is thus expected
unless the shock is hidden under the cooler wind.
P2 is due to the steepening of last bits of excess matter on the disk
which is delayed by the viscous time-scale. 
The viscous time which a ring of matter takes after it leaves the
Keplerian disk from a transition radius ($r_{tr}$, where the flow deviates from 
a Keplerian disk (Chakrabarti \& Titarchuk, 1995), 
and enters the post-shock region is given by
$t_{visc} \sim \frac{1}{\alpha}(\frac{h}{r})^{-2}\frac{r}{v_{Kep}}=10(\frac{0.01}{\alpha})
(\frac{0.1}{c_s})^2(\frac{r_{tr}}{100})^{1/2}$s where $\alpha$ is the
Shakura-Sunyaev viscosity parameter, $h(r)\sim c_s r^{3/2}$ is the dimensionless instantaneous
height of the disk (in vertical equilibrium) at a radius $r$ (measured in units of 
$R_g=2GM/c^2$, the Schwarzschild radius), $v_{Kep}$ is the rotational velocity
of the Keplerian orbit, and $c_s$ is the speed of sound
in units of velocity of light. As this ring of matter propagates through
this region, it is illuminated by soft photons coming out of the Keplerian disk, but since it 
is outside the shock, its radiations do not participate in the oscillation.
On the other hand, since rising side of P1 is in Off state, it shows QPO. 
Due to Compton cooling the spectrum of P1 is softer. However P2 occurs when 
the excess matter is almost entirely drained out from the disk. Hence its spectrum is harder.

It is to be noted that whether or not shocks would form depend on viscosity. It has been shown
(Chakrabarti, 1990; Chakrabarti \& Molteni, 1995; Chakrabarti 1996) that there is a critical viscosity
parameter $\alpha_c$ below which shocks can form and this parameter is about $\alpha_c \sim 0.015$. 
Beyond this chosks disappear and Keplerian flows directly enter into the black hole.
Out choice of $\alpha=0.01$ to explain the time scale is thus consistent with the presence of shocks.

In general, as excess matter drains out of the CENBOL, its optical depth decreases
and the spectrum gets harder in the On$^{++}$ state. This is observed in both RXTE and IXAE data.

\subsection{Origin of $\rho$ type light curve}

One of the reasons we plotted Fig. 5 using the particular
region of $\kappa$ class observations is that it may hold the
key of understanding the general light curves. The Figure clearly indicates
that $\rho$ type regions are peeled off gradually in On$^{++}$ states
of $\kappa$ class light curve. Arrows in Fig. 5 indicate that the forms
of rise and fall are qualitatively similar, but quantitatively one
is a miniature version of a fully developed $\rho$ type of bursts.
Details of the modeling is being discussed elsewhere (Chakrabarti et al. 2000b).

Briefly, when the entire CENBOL behaves like a single blob,  
it's light curve is of  $\rho$ type: count rate going up in Comptonization
time scale and rapidly drops in infall time-scale. However, in presence
of strong winds, and subsequent return of matter on the disk, turbulence are generated
and one may imagine that each of the turbulent cells after being steepened into small shocks,
producing mini-$\rho$ light curves, depending on the shapes and sizes of the turbulent cells.  
Each mini-shock produces a mini-$\rho$ curve after filing smaller sonic surface.

\section{Concluding remarks}

In this Paper, we have pointed out some new and interesting behaviour of the
light-curve of the black hole candidate  GRS 1915+105. We showed that very often the On states
of $\rho$ and $\nu$ types of light curves produce more than one peak 
which  behave differently. For instance, the first peak
may show QPOs whereas the second peak shows no QPO. 
The second peak also has a harder spectrum. In the $\kappa$ and $\lambda$ classes
when the On state duration is longer, the On state shows very noisy and oscillatory
behaviour towards the end, which we term as On$^{++}$ state. What is more,
photons from upper-half of these light curves do not show QPOs while those from the lower-half do.

Though these observations are related to very small regions of the overall light curves,
they  help us understand the behaviour of matter close to a black hole. If one assumes that
the duration of the Off states are determined by the time in which the optical depth of the
sonic sphere becomes unity, as CM00 suggested, then many puzzles are resolved.
In this picture, part of the cooler matter of the outflowing wind falls back on the CENBOL and
the pre-shock disk and its drainage time gives rise to the On states. It is possible 
that the first catastrophic cooling gives rise to the
first hiccup in the On state and the last significant density perturbation  due to the fallen matter
may cause the final hiccup. This can explain why P1 often shows QPO, but P2 does not.
Duration of On states in between may vary, depending on drainage time, giving rise to
various classes of light curves. Also, towards the end of the drainage
period, i.e., in On$^{++}$ states, the excess
matter is depleted and the signs of QPO starts appearing. What is most interesting,
mini-Off states show QPO, while mini-On states do not, indicating that
states with broader On are possibly made of $\rho$ type bursts.

This work is partly supported by Indian Space Research Organization (ISRO) through a project
entitled Quasi Periodic Oscillations in Black Holes. AN's work is supported by a DST project
entitled Analytical and Numerical Studies of Astrophysical Flows Around Compact Objects.
We gratefully acknowledge using RXTE data from public archive of GSFC/NASA.

{}

\end{document}